\documentclass[10pt,english]{iopart}
\usepackage[T1]{fontenc}
\usepackage[latin9]{inputenc}
\usepackage{graphicx}
\usepackage{setspace}
\usepackage{amssymb}
\usepackage{iopams}
\makeatletter

\providecommand{\boldsymbol}[1]{\mbox{\boldmath $#1$}}

\usepackage{babel}
\makeatother
\begin{document}

\title{The EPR experiment in the energy-based stochastic reduction framework}
\author{J Silman$^1$, S Machnes$^1$, S Shnider$^2$, L P Horwitz$^{1,3}$ and A Belenkiy$^2$}

\address{$^1$School of Physics and Astronomy, Tel-Aviv University, Tel-Aviv, Israel}
\address{$^2$Department of Mathematics, Bar-Ilan University, Ramat-Gan, Israel}
\address{$^3$Department of Physics, Ariel University Center of Samaria, Ariel, Israel}

\begin{abstract}
\textbf{We consider the EPR experiment in the energy-based stochastic
reduction framework. A gedanken set up is constructed to model the
interaction of the particles with the measurement devices. 
The evolution of particles' density matrix is analytically derived. 
We compute the dependence of the disentanglement rate on the parameters of the model, 
and study the dependence of the outcome probabilities on the noise trajectories. 
Finally, we argue that these trajectories can be regarded as non-local hidden variables.}
\end{abstract}

\section{Introduction}

A pure quantum state of a system is a vector in a Hilbert space, which
may be represented as a linear combination of a basis of eigenstates
of an observable (self-adjoint operator) or of several commuting observables.
Let us suppose that the eigenvalues corresponding to the eigenstates
of the Hamiltonian operator of a system are the physical quantities
measured in an experiment. If the action of the experiment is modeled
by a dynamical interaction induced by a term in the Hamiltonian of
the system, and its effect is computed by means of the standard evolution
according to the Schr\"{o}dinger equation, the final state would
retain the structure of the original linear superposition. One observes,
however, that the experiment provides a final state that is one of
the basis eigenstates and the superposition has been destroyed. The
resulting process is called reduction or collapse of the wave function.

The history of attempts to find a systematic framework for the description
of this process goes back very far in the development of quantum theory
(e.g., the problem of Schr\"{o}dinger's cat \cite{Schrodinger}).
In recent years significant progress has been made. Rather than invoking
some random interaction with the environment and attributing the observed
decoherence, i.e. collapse of a linear superposition, to the onset
of some uncontrollable phase relation, more rigorous methods have
been developed. These methods add to the Schr\"{o}dinger equation
stochastic terms corresponding to Brownian fluctuations of the wave
function \cite{Gisin0, Diosi, Gisin, GPR, Percival, Hughston, Horwitz, Adler},
generally understood as arising from the presence of the measurement
device.\\

In this paper, we apply some of these state reduction methods to the
phenomena considered in Bohm's formulation \cite{Bohm} of the Einstein-Podolsky-Rosen
paradox \cite{EPR} (henceforth the EPRB paradox), later analyzed
by Bell for its profound implications \cite{Bell}, and explored experimentally
by Aspect et al. \cite{Aspect}. The system to be studied consists
of a pair of spin-$\frac{1}{2}$ particles in the singlet state \begin{equation}
|\psi_{s}\rangle=\frac{1}{\sqrt{2}}(|\uparrow\downarrow\rangle-|\downarrow\uparrow\rangle)\,,\label{singlet}\end{equation}
the arrows denoting the spin components of the particles relative
to some arbitrary axis. The determination of the spin state of one
of the particles implies with certainty the spin state of the other,
even when the particles are very far apart. The particles are therefore
said to be entangled.

The question is often raised as to how the state of the second particle
can respond to the arbitrary choice of direction in the measurement
of the first. This question is dealt with here by the construction
of a gedanken set up describing the interaction of the particles with
the measurement devices. On this basis, using the mathematical
models recently developed for describing the reduction, or collapse,
of the wave function, we answer this question and give a mathematical
description of the process underlying such a measurement.\\

The paper is organized as follows. We begin in section 2 by reviewing the energy-based stochastic extension
of the one-particle Schr\"{o}dinger equation, and discuss its generalization to noninteracting multiparticle systems.
In section 3 we present a gedanken set up for studying the EPR experiment and show that it leads to the expected quantum mechanical predictions.
Next, in section 4 we analytically compute the stochastic expectation of the particles' density matrix, and quantify their
disentanglement rate. In section 5 we simulate the evolution of
the state of the particles for different random realizations of the noise, and argue that the noise trajectories can
be regarded as nonlocal hidden-variables. We end by discussing future avenues of research.
  
\section{\label{sec:Review}Energy-based stochastic state reduction}

\subsection{Energy-based stochastic extension of the Schr\"{o}dinger equation}

In the energy-based stochastic reduction framework the Schr\"{o}dinger
equation is extended as follows \cite{Hughston, Horwitz, Adler} \begin{eqnarray}
d\left|\psi(t)\right\rangle & = & -i\hat{H}\left|\psi(t)\right\rangle dt-\frac{\varsigma^{2}}{8}(\hat{H}-H(t))^{2}\left|\psi(t)\right\rangle dt\nonumber \\ & & +\frac{\varsigma}{2}(\hat{H}-H(t))\left|\psi(t)\right\rangle dW(t)\,. 
\label{Hughston}\end{eqnarray}
Here $H(t)\hat{=}\langle\psi(t)|\hat{H}|\psi(t)\rangle$ (we assume throughout
that $\left|\psi(t)\right\rangle $ is normalized, in consistence with eq. (\ref{Hughston})) as it is norm preserving, $W(t)$ is a standard
Wiener process, and $\varsigma$ is a parameter characterizing the reduction time scale the rate. 
(Note the choice of "natural" units $\hbar=c=1$.
 Accordingly, all quantities throughout the paper are expressed in units of length $[\ell]$.)

From the It\^{o} calculus rules it immediately follows that the above
process has two basic properties 

\begin{enumerate}
\item Conservation of energy \begin{equation}
H(t)=H(0)+\varsigma\int_{0}^{t}dW(s)V(s)\,,\label{energy process}\end{equation}
where $V(t)\hat{=}\langle\psi(t)|(\hat{H}-H(t))^{2}|\psi(t)\rangle$ is
the variance of the energy process $H(t)$.
\item Stochastic reduction \begin{equation}
dV(t)=-\varsigma^2 V^{2}(t)dt+\varsigma\beta(t)dW(t)\,,\label{variance process}\end{equation}
 where $\beta(t)\hat{=}\langle\psi(t)|(\hat{H}-H(t))^{3}|\psi(t)\rangle$
is the third moment of the energy deviation.
\end{enumerate}
It follows from eq. (\ref{variance process}) that the expectation
$E[V(t)]$ of the variance process obeys the relation \cite{Hughston,Horwitz}
\begin{equation}
E[V(t)]\leq E[V(0)]-\varsigma^2 \int_{0}^{t}dsE[V(s)]^{2}\,.\label{variance expectation}\end{equation}
Since $V(t)$ is positive, this implies that $E[V(t\rightarrow\infty)]\rightarrow0$
and (up to measure zero fluctuations) $V(t\rightarrow\infty)\rightarrow0$.
And since $V(t)=\langle\psi(t)|(\hat{H}-H(t))^{2}|\psi(t)\rangle$,
$V(t)=0$ implies $\langle\psi(t)|(\hat{H}-H(t))|\psi(t)\rangle=0$
or $\hat{H}\left|\psi(t)\right\rangle =H(t)\left|\psi(t)\right\rangle $,
so that $\left|\psi(t)\right\rangle $ is an eigenstate of the Hamiltonian.
Assuming no degeneracy, the system therefore reduces to one or another
of the eigenstates of the Hamiltonian $\hat{H}$, in accordance with
the statistical predictions of standard quantum mechanics \cite{Horwitz}.
Therefore, the expectation of the final configuration 
$E[\left|\psi(t\rightarrow\infty)\right\rangle \left\langle \psi(t\rightarrow\infty)\right|]$
corresponds to a mixed state, with each of the diagonal elements an
eigenstate of $\hat{H}$.

Note that the framework we have described cannot differentiate
between degenerate eigenstates. When this is the case, as in the standard
theory \cite{Luders}, the reduction process drives the system to
degenerate subspaces with the original relative phase between the
spanning eigenstates remaining unchanged.

\subsection{Extension to noninteracting multiparticle systems}

Nothing in the previous subsection limits the discussion to single
particle systems. The Hamiltonian in eq. (\ref{Hughston}) may just
as well represent a multiparticle system. This, however, is not the
only possible generalization to multiparticle systems, and indeed
there are cases where it is not suitable. To see this, and in anticipation
of the next section, let us consider a pair of noninteracting particles
$A$ and $B$. The Hamiltonian is now a direct sum \begin{equation}
\hat{H}=\hat{H}_{A}\oplus\hat{H}_{B}\,.\label{H direct sum}\end{equation}
We assume that the particles are very far apart, and that the environment
does not carry pervasive long-range correlations. Under these conditions the evolution of the
state's stochastic expectation $E[\rho (t)]$ ($\rho(t)=\left|\psi(t)\right\rangle \left\langle \psi(t)\right|$) should be local, in the sense that no correlations, quantum or classical, are generated.

Bearing this in mind, let us plug the Hamiltonian eq. (\ref{H direct sum})  
into eq. (\ref{Hughston}). Averaging over the noise we
 obtain a Lindblad type equation \cite{Sudarshan,Lindblad} for the state's stochastic expectation 
\begin{equation}
\frac{d}{dt}E[\rho(t)]=-i[\hat{H},\, E[\rho(t)]]-\frac{\varsigma^{2}}{8}\sum_{i,\, j=A,\, B}[\hat{H}_{i},\,[\hat{H}_{j},\, E[\rho(t)]]]\,.\label{nonlocal Lindblad}\end{equation}
This equation is causal (does not allow for superluminal signalling), as is
easily established by tracing over any of the two subsystems. However, the mutual information\footnote{The mutual information of two systems is defined as $I_{AB}=S_{A}+S_{B}-S_{AB}$, where $S_{i}$ and $S_{AB}$ are the von-Neumann entropies of system $i$ and the composite system, respectively, and serves as a quantitative measure of the total
amount of correlations, quantum and classical, between the systems.} may increase with time. The evolution is therefore nonlocal, as may well have been expected considering 
that both systems are driven by the same noise.

However, a local evolution equation for the state's stochastic expectation can be
achieved if we have each of the systems driven by an independent noise term.
This means that eq. (\ref{Hughston}) must be generalized as \begin{eqnarray}
d\left|\psi(t)\right\rangle  & = & -i\hat{H}\left|\psi(t)\right\rangle dt-\frac{1}{8}\sum_{i=A,\, B}{\varsigma_{i}}^{2}(\hat{H}_{i}-H_{i}(t))^{2}\left|\psi(t)\right\rangle dt\nonumber \\
 &  & +\frac{1}{2}\sum_{i=A,\, B}{\varsigma_{i}}(\hat{H}_{i}-H_{i}(t))\left|\psi(t)\right\rangle dW_{i}(t)\label{generalization}\end{eqnarray}
where $\varsigma_{i}$ governs the reduction rate of particle $i$ to the eigenstates
 of $\hat{H}_{i}$ and $dW_{i}(t)dW_{j}(t)=\delta_{ij}dt$. Indeed,
the above process invariably drives the system to product states of
the form $\left|E_{A}\right\rangle \otimes\left|E_{B}\right\rangle$,
where $\hat{H}_{i}\left|E_{i}\right\rangle =E_{i}\left|E_{i}\right\rangle$,
with the same probabilities as predicted by the standard theory \cite{Horwitz}.
The corresponding Lindblad equation for the state's stochastic expectation is now given by 
\begin{equation}\frac{d}{dt}E[\rho(t)]=-i[\hat{H},\, E[\rho(t)]]-\sum_{i=1,\,2}\frac{{\varsigma_{i}}^2}{8}[\hat{H}_{i},\,[\hat{H}_{i},\, E[\rho(t)]]]\,. \label{Lindblad}\end{equation} Note that in standard quantum theory this
evolution can only arise from the respective coupling of a pair separate noninteracting systems to noncorrelated environments. 

\section{Stochastic reduction in the EPRB experiment}

In this section we construct a gedanken set up to show how the energy-based
stochastic reduction framework can provide us with a consistent description
of the EPRB experiment.

We consider a pair of spin-half particles, $A$ and $B$, with vanishing
total spin and momentum, moving in opposite directions. Along the path
of each particle a spin measurement device is placed. With no loss
of generality we assume that the measurement device in the path
of particle $A$ measures its spin component along $\hat{z}$, and that
the measurement device in the path of particle $B$ measures its spin component along
$\hat{n}$, where $\hat{n}$ is some unit vector pointing in an arbitrary
direction.

So long as the particles are far from the measurement devices
the Hamiltonian governing their (free) evolution is given by \begin{equation}
\hat{H}_{0}=\frac{\hat{p}_{A}^{2}}{2m_{A}}+\frac{\hat{p}_{B}^{2}}{2m_{B}}\,.\label{free hamiltonian}\end{equation}
However, once the particles approach some neighborhood of the detectors
(which we assume happens simultaneously) we assume that $\hat{H}_0$ 
is corrected by the addition of a perturbation  \begin{equation}
\hat{H}_{int}=\mu_{A}\sigma_{A}^{z}\otimes\boldsymbol{1}+\mu_{B}\boldsymbol{1}\otimes\sigma_{B}^{\hat{n}}\label{interaction hamiltonian}\end{equation}
describing the \emph{local} interaction of the particles with the
measurement devices. Here $\mu_{i}$ denotes the strength of the
coupling of particle $i$ to the corresponding measurement device and $\sigma_{B}^{\hat{n}}=\boldsymbol{\sigma}\cdot\hat{n}$.
The eigenstates of the perturbed Hamiltonian $\hat{H}=\hat{H}_{0}+\hat{H}_{int}$
are products of momentum and spin eigenstates, and are fully specified
by the four eigenvalues $-\infty<p_{A},\, p_{B}<\infty$ and $\sigma_{A}^{z},\,\sigma_{B}^{\hat{n}}=\pm1$.

The continuous spectrum of the momentum operators gives
rise to an irremovable degeneracy in $\hat{H}$. Nevertheless, for
wave packets localized in momentum space and sufficiently large values
of the $\mu_{i}$ this residual degeneracy is negligible. In all that follows we shall therefore make the
 approximation \begin{equation}
\hat{H}\simeq\hat{H}_{int}\,.\label{approx}\end{equation} 
The four possible spin outcomes are then given by $\left|\uparrow\nearrow\right\rangle $,
$\left|\uparrow\swarrow\right\rangle $, $\left|\downarrow\nearrow\right\rangle $,
and $\left|\downarrow\swarrow\right\rangle $, the slanted (vertical)
up and down arrows denoting spin-up and spin-down eigenstates of $\sigma_{B}^{\hat{n}}$
($\sigma_{A}^{z}$), respectively, with corresponding probabilities
\begin{eqnarray}
& & P_{\,\uparrow\swarrow}=P_{\,\downarrow\nearrow}=\frac{1}{2}\cos^{2}\frac{\theta}{2}\,,\qquad P_{\,\uparrow\nearrow}=P_{\,\downarrow\swarrow}=\frac{1}{2}\sin^{2}\frac{\theta}{2}\,,\nonumber \\
& & \label{outcome prob}\end{eqnarray}
where $\theta$ is the angle between the $\hat{z}$ and $\hat{n}$
axes. The reduction process ultimately reproduces the results of the standard
theory.

It is important, lending further credibility to our set up, that 
the measurement of the spin of only one of the particles suffices to induce
the reduction. This is clearly seen by recasting $\hat{H}_{int}$
in its diagonal form \begin{equation}
\hat{H}_{int}=\sum_{i=\uparrow,\,\downarrow}\sum_{j=\nearrow,\,\swarrow}\lambda_{ij}\left|ij\right\rangle \left\langle ij\right|\,,\label{interaction  Hamiltonian diagonalized}\end{equation}
where $\lambda_{\uparrow\nearrow}=-\lambda_{\downarrow\swarrow}=\ \mu_{A}+\mu_{B}$
and $\lambda_{\uparrow\swarrow}=-\lambda_{\downarrow\nearrow}=\mu_{A}-\mu_{B}$.
Indeed, this is approximately what will be observed at a time $t\simeq 1/\varsigma_{i}$ 
when $\varsigma_{i}\ll \varsigma_{j\neq i}$.

We also note that when $\mu_{A}$ equals $\mu_{B}$ ($-\mu_{B}$) the degeneracy
of the states $\left|\uparrow\swarrow\right\rangle $ and $\left|\downarrow\nearrow\right\rangle $
($\left|\uparrow\nearrow\right\rangle $ and $\left|\downarrow\swarrow\right\rangle $)
is not removed. This, however, is merely an artifact of not taking
into account the quantum nature of the measurement devices and
the fields generating the coupling. Indeed, the \textit{local} uncertainty of these
fields, together with the absence of significant long-range correlations, serves to lift this degeneracy. 

Finally, we should remark that if the two particles are identical,
then the Hamiltonian must be symmetric under the interchange of the
particle indices. However, since the particles are very far apart
when the measurement takes place, there is no overlap of the one particle
wave functions, and the symmetrization or anti-symmetrization of
the composite wave function is not required. Thus, the presence of
two widely separated measurement devices can split the degeneracy
into distinct states, which can, in fact imply that $\hat{H}_{int}$
is not symmetric under particle exchange.

\section{Evolution of the state's stochastic expectation}

To study the evolution of the state's stochastic expectation in our set up we 
simply substitute $\mu_{A}\sigma_{A}^{z}$ and $\mu_{B}\sigma_{B}^{\hat{n}}$ for 
$\hat{H}_{A}$ and $\hat{H}_{B}$ in eq. (\ref{Lindblad}). We thus have
\begin{eqnarray}
 \frac{d}{dt}E[\rho(t)] & = & -i\mu_{A}[\sigma_{A}^{z},\, E[\rho(t)]]-\frac{\varsigma_{A}^{2}\mu_{A}}{8}[\sigma_{A}^{z},\,[\sigma_{A}^{z},\, E[\rho(t)]]]\nonumber \\
& &  -i\mu_{B}[\sigma_{B}^{\hat{n}},\, E[\rho(t)]]-\frac{\varsigma_{B}^{2}\mu_{B}}{8}[\sigma_{B}^{\hat{n}},\,[\sigma_{B}^{\hat{n}},\, E[\rho(t)]]]\,.\label{Lindblad specific}\end{eqnarray}
This equation is linear in $E[\rho(t)]$, and can therefore be transformed into
a linear equation for a vector whose components are the sixteen elements
of $E[\rho(t)]$. An analytical solution can then be obtained by bringing the $16\times 16$ matrix,
representing the action of the operators on the right-hand side of the equation, to its Jordan normal form. When
working in the basis {$\uparrow\nearrow$, $\uparrow\swarrow$, $\downarrow\nearrow$,
$\downarrow\swarrow$} the solution reads as follows 

 \begin{eqnarray}
&  & \frac{1}{2}\left(\begin{array}{cccc}
\sin^{2}\frac{\theta}{2} & -\frac{1}{2}Q_{B}^{\star}\sin\theta & \frac{1}{2}Q_{A}^{\star}\sin\theta & Q_{A}^{\star}Q_{B}^{\star}\sin^{2}\frac{\theta}{2}\\
-\frac{1}{2}Q_{B}\sin\theta & \cos^{2}\frac{\theta}{2} & -Q_{A}^{\star}Q_{B}\cos^{2}\frac{\theta}{2} & -\frac{1}{2}Q_{A}^{\star}\sin\theta\\
\frac{1}{2}Q_{A}\sin\theta & -Q_{A}Q_{B}^{\star}\cos^{2}\frac{\theta}{2} & \cos^{2}\frac{\theta}{2} & \frac{1}{2}Q_{B}^{\star}\sin\theta\\
Q_{A}Q_{B}\sin^{2}\frac{\theta}{2} & -\frac{1}{2}Q_{A}\sin\theta & \frac{1}{2}Q_{B}\sin\theta & \sin^{2}\frac{\theta}{2} \end{array}\right) \nonumber \\
& & \label{DM solution} \end{eqnarray}
with $Q_{i}\hat{=}\exp\left(\mu_{i}\left(4i-\mu_{i}\varsigma_{i}^{2}\right)t/2\right)$
($i=A,\, B$), and $\theta$ the angle between $\hat{z}$ and $\hat{n}$. We note that the $Q_{i}$ can be decomposed into a product of 
$\exp\left(2i\mu_{i}t\right)$ and $\exp\left(-\mu_{i}^2\varsigma_{i}^{2}t/2\right)$ representing the contibutions of the unitary and 
stochastic processes, respectively.
We also note that in the limit that $t\rightarrow\infty$ the off-diagonal
terms vanish and the density matrix reproduces the expected measurement
outcomes and probabilities.

\begin{figure}[!t]
\centering
\begin{minipage}[htbp]{0.8\linewidth}
\centering
\includegraphics[width=0.8\linewidth]{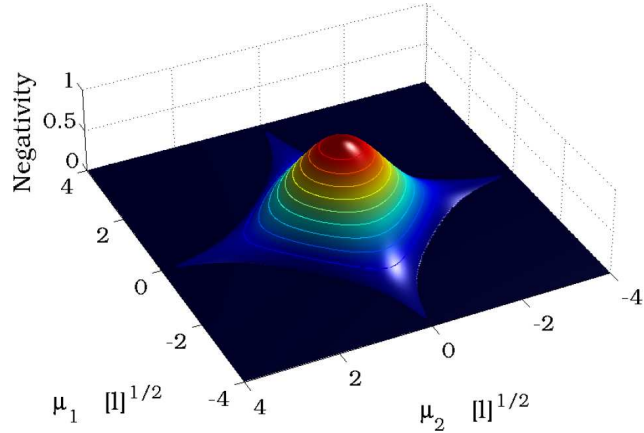}
\caption{(Color online) The negativity as a function of the coupling strengths $\mu_{A}$
and $\mu_{B}$. ($\varsigma_{A}=\varsigma_{B}=1\,[\ell]^{1/2}$, $t=1\,[\ell]$, and
$\theta=3\pi/5$) On the (dark blue) flat surface the negativity equals
zero. The disentanglement time is finite unless $\mu_{A}$ or $\mu_{B}$
vanishes.}
\end{minipage}
\end{figure}

We now wish to examine the transition of the stochastic expectation
of the density matrix from the initial maximally entangled singlet
state to the mixture of the final outcomes, paying particular attention
to the rate of the disentanglement. While for pure states, the entanglement
is quantified by the von-Neumann entropy, there is no single measure of entanglement
for mixed states. One of the standard measures is the negativity \cite{negativity},
$\frac{1}{2}(\left\Vert \rho^{T_{i}}\right\Vert -1)$, where $\rho^{T_{i}}$, the partial 
transpose with respect to $i$ of $\rho$,
is obtained from $\rho$ by transposing the indices of system $i$%
\footnote{When considering the entanglement between two systems it is irrelevant
which of the indices is transposed.}, 
which is just minus the sum of the negative eigenvalues of the $\rho^{T_{i}}$ \cite{peres,Horodecki}.

\begin{figure}[!t]
\centering
\begin{minipage}[t]{0.474\linewidth}
\centering
\includegraphics[width=1.0\linewidth]{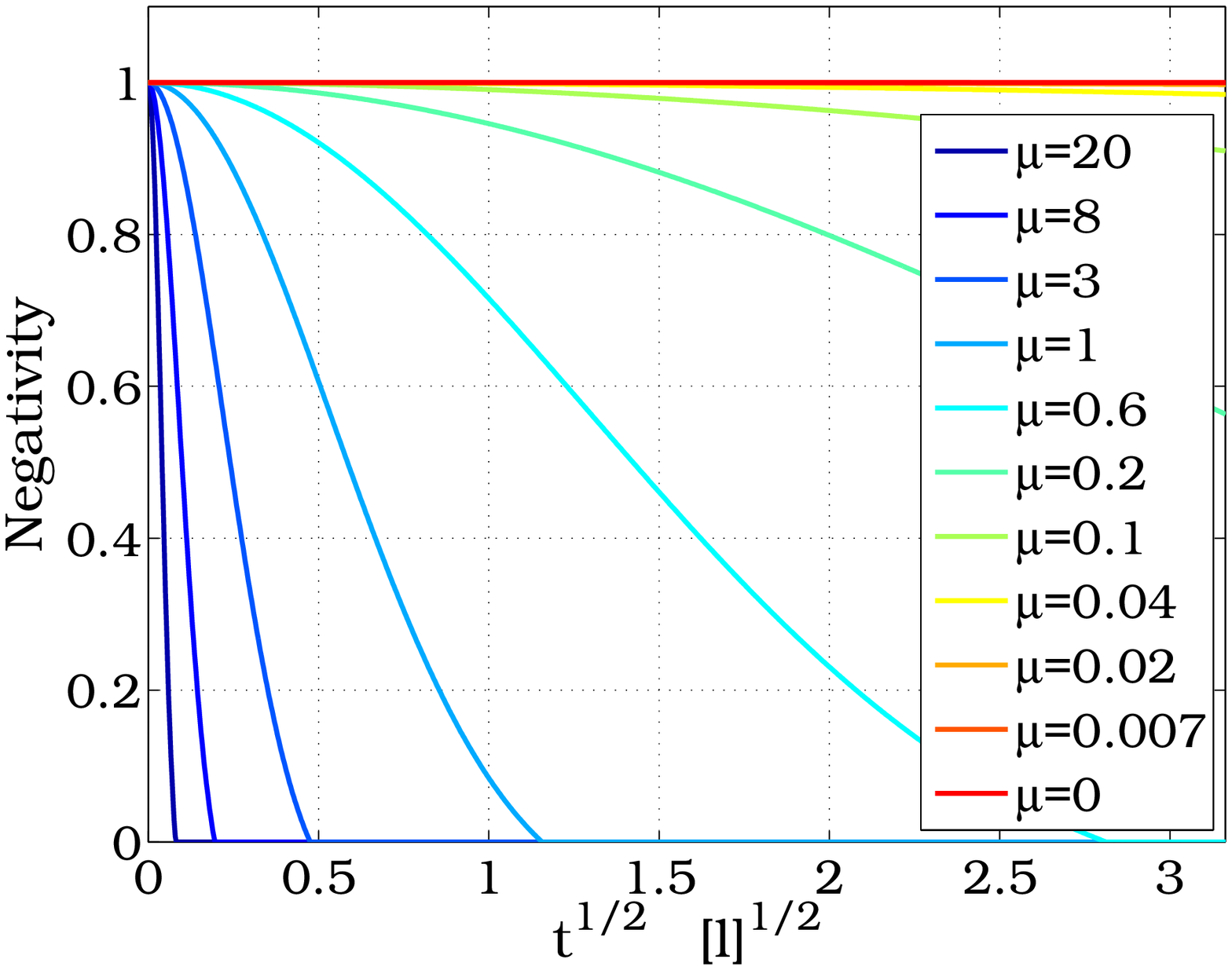}
\caption{(Color online) The negativity as function of time for different values of the
 $\varsigma_{A}= \varsigma_{B}$. ($\theta=3\pi/5$, $\mu_{A}=\mu_{B}=1\,[\ell]^{-1}$) 
The legend gives the value of ${\varsigma_i}^2/8$ for each of the curves.} \label{fig:side:a}
\end{minipage}%
\hspace{0.05\linewidth}%
\begin{minipage}[t]{0.474\linewidth}
\centering
\includegraphics[width=1.0\linewidth]{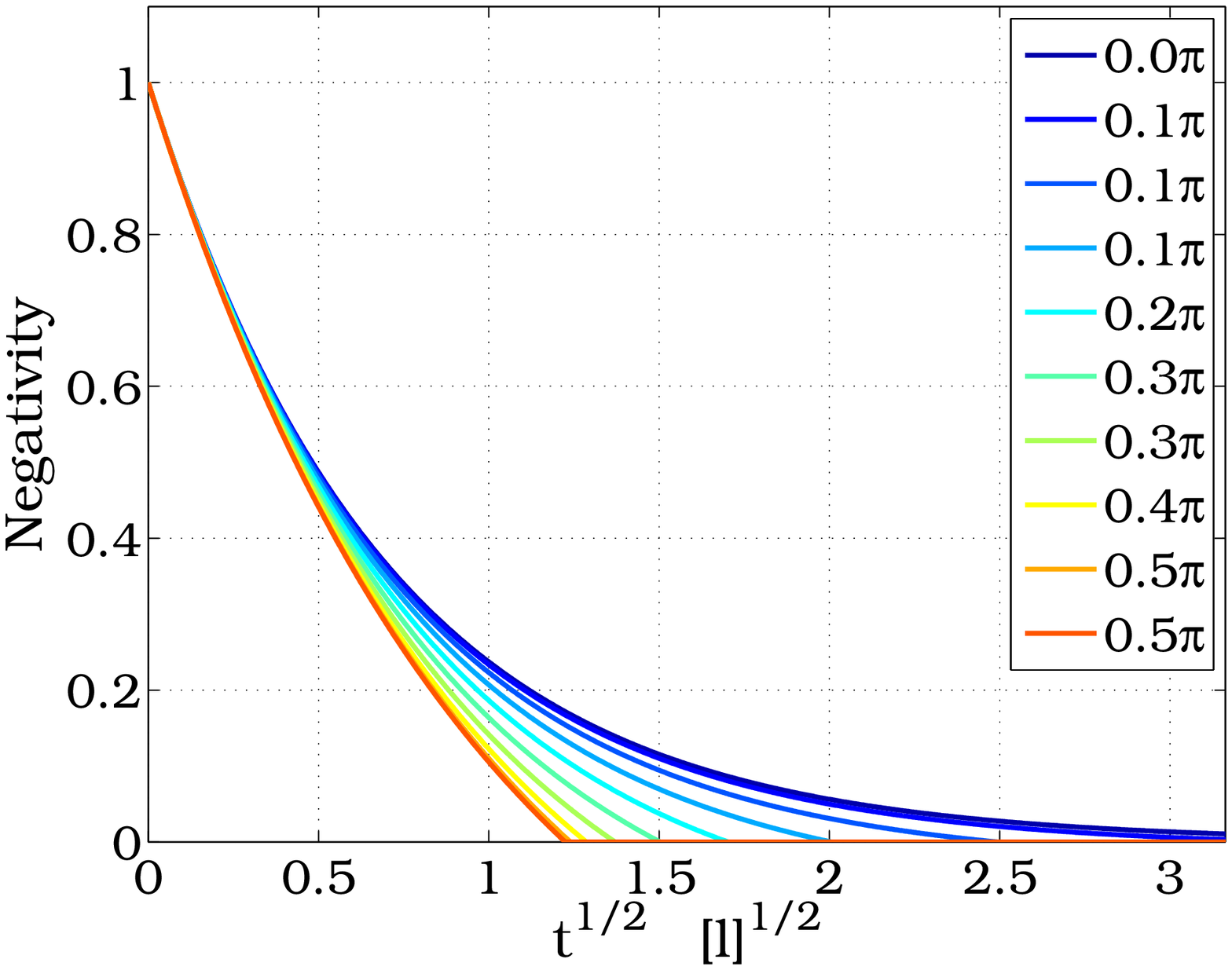}
\caption{(Color online) The negativity as function of time and 
$\theta$. ($\varsigma_{A}=\varsigma_{B}=1\, [\ell]^{1/2}$, $\mu_{A}=\mu_{B}=1$) 
The legend gives the value of $\theta$ for each of the curves.}\label{fig:side:b}
\end{minipage}
\end{figure}

Numerical results are presented in figures (1-3). Figure (1) illustrates
the dependence of the negativity on the coupling strengths, $\mu_{A}$
and $\mu_{B}$. We see that unless $\mu_{A}$ or $\mu_{B}$ vanishes, the disentanglement
time is finite. This phenomenon, termed "entanglement sudden death"  \cite{Eberly}, is not unique to our setting and is typical of open systems dynamics \cite{Halliwell}.
 For short times the rate of disentanglement is roughly dependent on 
$\sqrt{{\mu_{A}}^2+{\mu_{B}}^2}$, while for long times becomes linear. 
Figures (2) and (3) show the negativity as function of
 time for different values of $\varsigma_A=\varsigma_B$
and $\theta$, respectively.%
\\


\section{Noise trajectories}

Eq. (\ref{generalization}) gives rise to an entropy conserving evolution. 
In particular, this means that a pure state remains pure. 
The trajectories realized by the $W_{i}(t)$
during the evolution fully specify the state's history. However, as
there is no means of determining these, all accessible
information is contained in the state's stochastic expectation. The
trajectories can therefore be regarded as hidden-variables\footnote{For
 a somewhat different approach see \cite{Hughston2006}.
}, i.e. indeterminable variables carrying information regarding the state of the system unavailable 
in the standard quantum mechanical description.

\begin{figure}[htbp]
\centering
\includegraphics[width=0.75\linewidth]{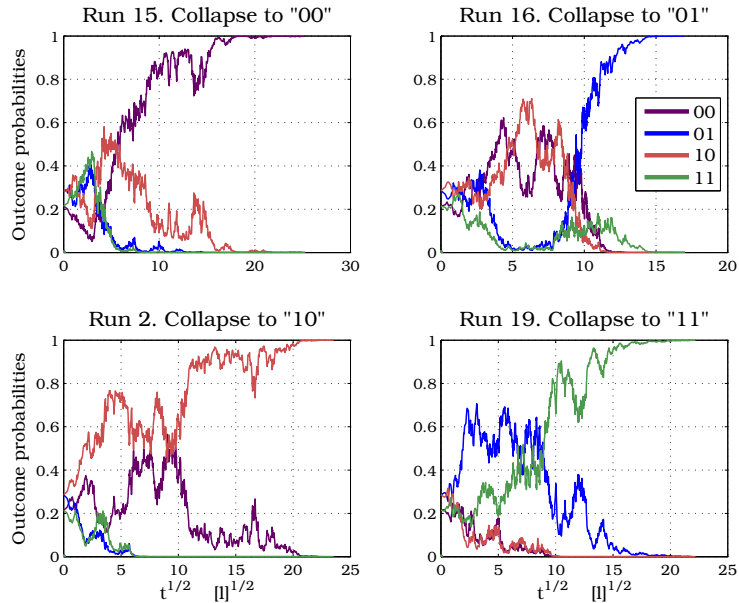}
\caption{(Color online) Four realizations of the collapse process. 
The probabilities for the four different outcomes ($1\hat{=}\uparrow$, $0\hat{=}\downarrow$) 
are plotted as a function of time.
$\mu_{A}=\mu_{B}=1\,[\ell]^{-1}$, $\varsigma_A=\varsigma_B=0.14\,[\ell]^{1/2}$ and $\theta=5\pi/{11}$. 
The same realization of $W_{A}(t)$ was used in all of the runs, whereas  $W_{B}(t)$ varies from one run to the next. We see that
the final outcome of particle $A$ depends on not only $W_{A}(t)$ but on $W_{B}(t)$ as well.}%
\end{figure}

From the relation $dW_{i}(t)dW_{j}(t)=\delta_{ij}dt$ it does not follow
that the trajectories are \emph{local} hidden-variables. Indeed, any
hidden-variable theory adhering to the statistical predictions of
standard quantum mechanics must violate some Bell inequality, and
as such our set up is manifestly nonlocal. Explicitly, this just
means that the \emph{both} $W_{A}(t)$ and $W_{B}(t)$
determine the final state of each of the particles (rather than $W_{i}(t)$ 
determining the final state of particle $i$ on its own). This point is illustrated 
in figures (4) and (5), which present the results of numerical
iterative solutions to eq. (\ref{generalization}) with randomly generated noise terms. Figure (4) explicitly
shows how, for the same realization of $W_{A}(t)$, different realizations
of $W_{B}(t)$ lead to different outcomes in the spin measurement
of particle $A$ (and particle $B$). It is also interesting that
in this case the probabilities for the measurement outcomes no longer
agree with those of quantum mechanics, as is evident from figure (5).

To see how this comes about we must go back to eq. (\ref{generalization}).
Even though it describes a pair of uncoupled systems, it gives rise to a
 potentially disentangling (and entangling) evolution, because 
each of the ${H_i}(t)$ depends on the full state of the system (${H_i}(t)={tr_{j\neq i}}[\rho(t)\hat{H}_i]$), and therefore on ${W_{j\neq i}}(s\leq t)$. Note, however, that this nonlocality does
not allow for superluminal signalling since the trajectories are "hidden".

\begin{figure}[htbp]
\centering
\includegraphics[width=0.75\linewidth]{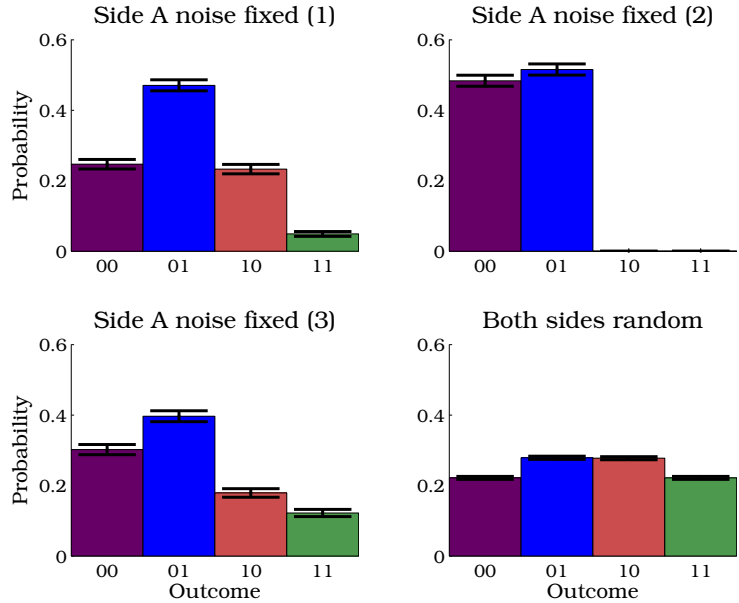}
\caption{(Color online) Dependence of the outcome probabilities on the noise trajectories. 
The first three figures display the results of 1000 runs simulations, in which $W_{A}(t)$ is held fixed from run to the next, but realized differently in each simulation, while $W_{B}(t)$  varies from one run to the next. 
We see that different realizations of $W_{A}(t)$ lead to different outcome probabilities 
that do not agree with those of quantum mechanics. In the fourth figure, displaying the results of 10000 runs simulation, both $W_{A}(t)$ 
and $W_{B}(t)$ vary, and the quantum mechanical predictions are obtained.}%
\end{figure}


\section{Some concluding remarks}

We have discussed Bohm's formulation of the EPR experiment in the
the energy-based stochastic reduction framework. In particular, we
have seen how the presence of the measurement devices induces
the reduction of the singlet state to the expected outcome product
states with correct probabilities as predicted by the standard theory 
and have given the explicit time evolution of the process of disentanglement.
As an extension of this idea, one may consider a problem with a natural
degeneracy of some initial state where the presence of effective detectors
of some type induces a perturbation in which stochastic reduction
takes place, as in the asymptotic cluster decomposition of products
of quantum fields reducing an $n$-body system to $m$ $k$-body systems,
or the formation of local correlations in $n$-body systems such as
liquids, or spontaneous symmetry breaking. In all these cases, due
to the existence of continuous spectra, there will be some residual
dispersion in the final state, although possibly very small. We are
currently studying possible applications of the methods discussed
here to such configurations.

\ack

J Silman and S Machnes thank B Reznik for invaluable discussions and acknowledge support
from the Israeli Science Foundation (Grant no. 784/06).

\end{document}